\documentclass[runningheads]{llncs}
\usepackage{graphicx}
\usepackage[absolute,overlay]{textpos}
\usepackage{enumitem}
\usepackage{amsmath}
\usepackage{textcomp}
\usepackage{float}
\newcommand*{\figref}[1]{Fig.~(\ref{fig:#1})}

\begin{document}
\title{Advanced Signal Reconstruction in Tunka-Rex with Matched Filtering and Deep Learning}
%
%
\author{
P.~Bezyazeekov\inst{1}
\and
N.~Budnev\inst{1}
\and
O.~Fedorov\inst{1}
\and
O.~Gress\inst{1}
\and
O.~Grishin\inst{1}
\and
A.~Haungs\inst{2}
\and
T.~Huege\inst{2,3}
\and
Y.~Kazarina\inst{1}
\and
M.~Kleifges\inst{4}
\and
D.~Kostunin\inst{5}
\and
E.~Korosteleva\inst{6}
\and
L.~Kuzmichev\inst{6}
\and
V.~Lenok\inst{2}
\and
N.~Lubsandorzhiev\inst{6}
\and
S.~Malakhov\inst{1}
\and
T.~Marshalkina\inst{1}
\and
R.~Monkhoev\inst{1}
\and
E.~Osipova\inst{6}
\and
A.~Pakhorukov\inst{1}
\and
L.~Pankov\inst{1}
\and
V.~Prosin\inst{6}
\and
F.~G.~Schr\"oder\inst{2,7}
\and
D.~Shipilov\inst{1}
\and
A.~Zagorodnikov\inst{1}
}%
\authorrunning{P. Bezyazeekov et al.}
%
%
%

\institute{
	Institute of Applied Physics ISU, Irkutsk, Russia
	\and 
	Institut f\"ur Kernphysik, KIT, Karlsruhe, Germany
	\and 
	Astrophysical Institute, Vrije Universiteit Brussel, Pleinlaan 2, Brussels, Belgium
	\and
	Institut f\"ur Prozessdatenverarbeitung und Elektronik, KIT, Karlsruhe, Germany
	\and
	DESY, Zeuthen, Germany
	\and
	Skobeltsyn Institute of Nuclear Physics MSU, Moscow, Russia
	\and
	Bartol Research Inst., Dept. of Phys. and Astron., Univ. of Delaware, Newark, USA
}
\maketitle              
\begin{abstract}
The Tunka Radio Extension (Tunka-Rex) is a digital antenna array operating in the frequency band of 30-80 MHz, measuring the radio emission of air-showers induced by ultra-high energy cosmic rays.
Tunka-Rex is co-located with the TAIGA experiment in Siberia and consists of 63 antennas, 57 of them in a densely instrumented area of about 1 km\textsuperscript{2}. 
The signals from the air showers are short pulses, which have a duration of tens of nanoseconds and are recorded in traces of about 5 \textmu s length.
The Tunka-Rex analysis of cosmic-ray events is based on the reconstruction of these signals, in particular, their positions in the traces and amplitudes.
This reconstruction suffers at low signal-to-noise ratios, i.e. when the recorded traces are dominated by background.
To lower the threshold of the detection and increase the efficiency, we apply advanced methods of signal reconstruction, namely matched filtering and deep neural networks with autoencoder architecture.
In the present work we show the comparison between the signal reconstructions obtained with these techniques, and give an example of the first reconstruction of the Tunka-Rex signals obtained with a deep neural networks.
\end{abstract}

\section{Introduction}

Cosmic rays (CR) are high-energy charged particles with extra-terrestrial origin.
The sources of ultra-high energy CR are still unknown due to complexity of tracing charged particles deflected by galactic and extragalactic magnetic fields.
However, the energy spectrum and mass composition of CR can shed light on the most powerful cosmic accelerators.
Due to low flux of ultra-high energy CR it is impossible to measure them directly (in space or higher layers of atmosphere), 
and they are detected by large ground detectors measuring cascades produced by their interaction with the atmosphere.
These cascades, called air-showers, consist of many secondary particles, including electrons and positrons, which produce short radio pulses due to deflection in the Earth's magnetic field.  
These pulses have a broadband spectrum mostly in the MHz domain and a duration of tens of nanoseconds~\cite{Schroder:2016hrv}.

\begin{figure}[t!]
	\centering
	\includegraphics[width=1\linewidth]{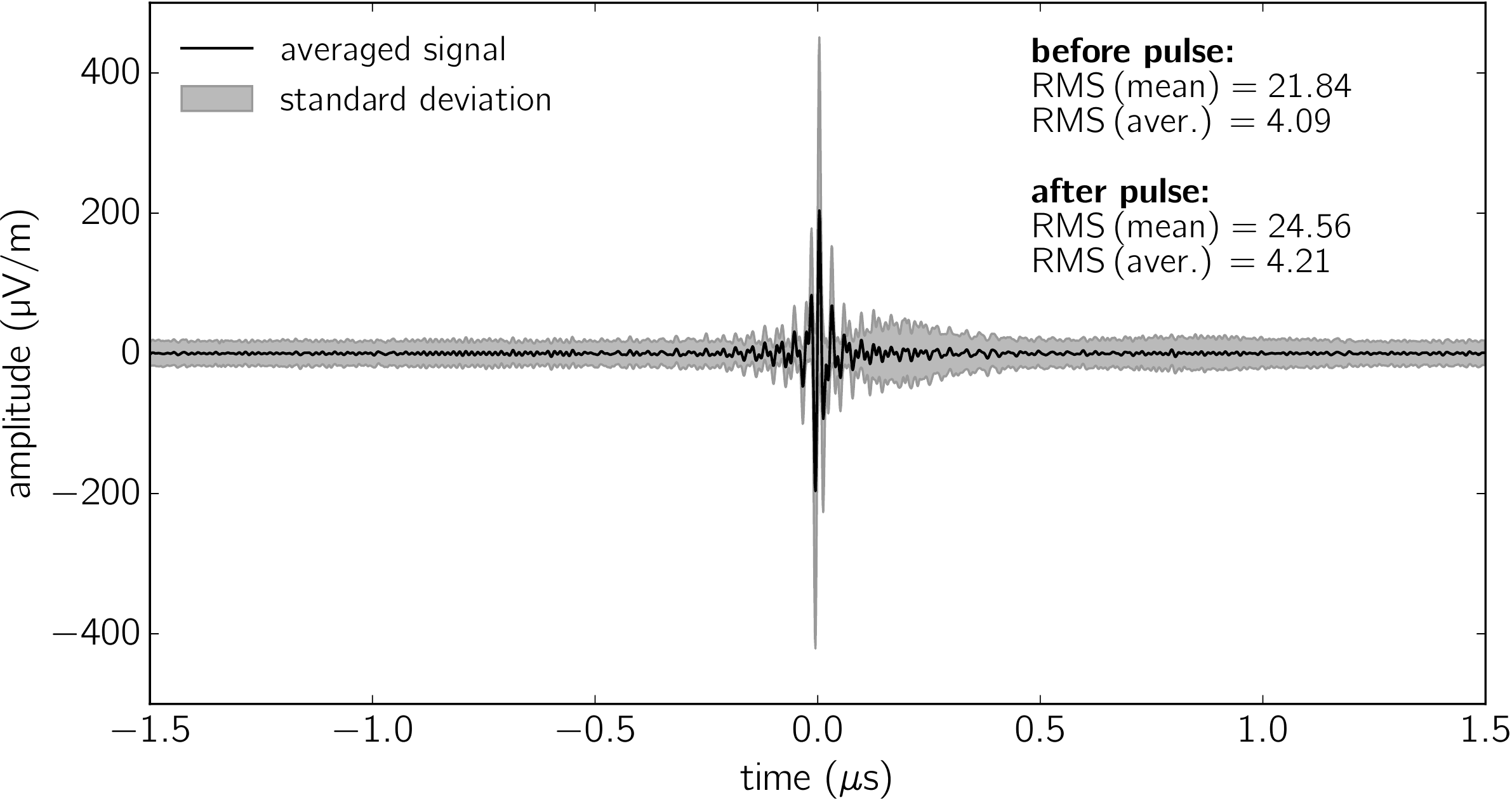}
	\\
	\caption{Average of 400 Tunka-Rex traces centered at the pulse. 
	 	The black line indicates the mean value, 
		the shaded area indicates one standard deviation.
		The expected reduction factor of the noise by averaging is ${\sqrt{400} = 20}$, however it is significantly less (about 5), moreover one can see prominent noise feature after the pulse.
	}
	\label{fig:nonwhite}
\end{figure}

Tunka-Rex~\cite{Bezyazeekov:2015rpa} is a sparse antenna array located at the TAIGA facility~\cite{Budnev:2017fyg,Kostunin:2019nzy} in the Tunka valley (Eastern Siberia).
It consists of 63 antennas measuring radio emission from air showers in the frequency band of 30-80 MHz.
Since Tunka-Rex is placed in a relatively radio-quiet location, the main background is from the Galaxy and has a power law behavior with an index of~$\approx-2.5$. 
However, there are plenty of non-stationary background sources, which complicate the reconstruction of events with low energies and may distort the reconstruction of the signal.
This background is caused by hardware RFI (especially after several upgrades of TAIGA facility) and by anthropogenic influence.
A simple study of the background shows that the noise in the Tunka-Rex traces has distinguishable features and cannot be approximated as white noise (see~\figref{nonwhite}).
In this work we discuss different approaches of signal reconstruction and background suppression, particularly a standard one used in many experiments and more sophisticated ones: \textit{matched filtering (MF)} and \textit{autoencoder (AE)}.
We discuss their performance and future applications.

\section{Signal reconstruction}

The air-shower pulse at each antenna station is measured in two independent channels and digitized in two traces of 1024 samples each with a rate of 200~MHz (i.e. binsize of 5 ns).
For the reconstruction of cosmic-ray air showers the two main properties of radio pulses are used: the amplitude of the signal and its arrival time.
Details of this reconstruction and its application to cosmic-ray science are given in Refs.~\cite{Bezyazeekov:2015ica,Bezyazeekov:2018yjw}.
In this paper we focus on the signal reconstruction itself and methods of its improvement.

Before the reconstruction of the signals we perform several preprocessing transformations.
Spectra of the signals obtained with Fourier transform are cut by a digital bandpass to 35-80~MHz and filtered with a median filter, 
which removes narrow-band RFI and equalizes the noise using a sliding window of 3~MHz width.
Then the traces are upsampled in order to increase the timing resolution by factors of 4, 16, and 64 for the standard method, AE, and MF, respectively.
As last step, the electric field along the two polarization directions in the plane perpendicular to the shower axis are reconstructed, namely $\mathbf{v}\times \mathbf{B}$ (along the Lorentz force, where $\mathbf{v}$ is the direction of the air shower and $\mathbf{B}$ the direction of the geomagnetic field) and $\mathbf{v}\times\mathbf{v}\times\mathbf{B}$ perpendicular to it.
Since the radio emission of air-showers is mostly in the $\mathbf{v}\times \mathbf{B}$ polarization we consider only this one hereafter.

The main quantity used for the definition of the threshold of signal reconstruction is the signal-to-noise ratio (SNR).
The SNR is defined using the following formula:
\begin{equation}
\mathrm{SNR} = {S}^{2}/{N}^{2}\,,
\end{equation}
where $S$ is the amplitude of the peak of the signal envelope ${S = \mathrm{max}(s(t))}$ inside of the \textit{signal window}, a 200~ns window defined by the trigger time.
The envelope $s(t)$ is defined as 
\begin{equation}
{s(t)} = u(t) + i\mathcal{H}[u(t)]\,,
\end{equation}
where $u(t)$ is the initial trace of the electric field at the antenna as a function of time $t$,
and $\mathcal{H}$ denotes the Hilbert transformation.
The noise level $N$ is defined as RMS in the \textit{noise window}.
$\mathrm{SNR}_{th}$ is defined as minimal RMS by sliding the noise window, a 500~ns window, scanning the trace in order to pass over occasional RFI.
We also use the additional definition of neighborhood $SNR_{neighb.}$ for the cases of overlapping signal with RFI (when noise is inside the signal window) based on the estimation of the noise level around the signal ($\pm$ 100 ns).
Thus, to keep the ratio of false positive detection at the level of 5\%, the thresholds are set as following: $\mathrm{SNR}_{th} \geq 16$ and $\mathrm{SNR}_{neighb.} \geq 10$.


\subsection{Signal and background dataset for simulation study}
In this study we use a dataset of 650\,000 samples of measured background (2014-2017) recorded by Tunka-Rex and 25\,000 CoREAS~\cite{Huege:2013vt} simulations.
The air-shower pulse is randomly located within the signal window, summed with noise and folded with the Tunka-Rex hardware response taking into account the geometry of the air shower and the calibration of the detector.

As was discussed in Ref.~\cite{Bezyazeekov:2018yjw}, the simulated signals reproduce real ones with satisfactory accuracy.
As shown below, the methods developed for simulated pulses can be applied to the real data without additional tuning.


\subsection{Matched filtering}
MF is a widely used technique of extracting signal from noisy data based on the convolution of a signal template $T$ with the time trace $u(t)$:
\begin{equation}
s_{cc}[t] = \sum_{t_0} T[t_{0}-t]u[t],~~A_{cc} = \mathrm{max}(s_{cc}[t_{0}])\,,
\end{equation}
where $A_{cc}$ is the maximum of the convolution, which defines the peak of the filtered signal $s_{cc}[t]$.

MF is very effective for the detection of the signal position inside a trace contaminated significantly by white noise and can even detect signals with ${\mathrm{SNR} \ll 1}$.
However, MF is unable to reconstruct the pulse shape and can only give an approximate estimation of the amplitude of initial signal.



\begin{figure}[t]
	\centering
	\includegraphics[width=0.48\linewidth]{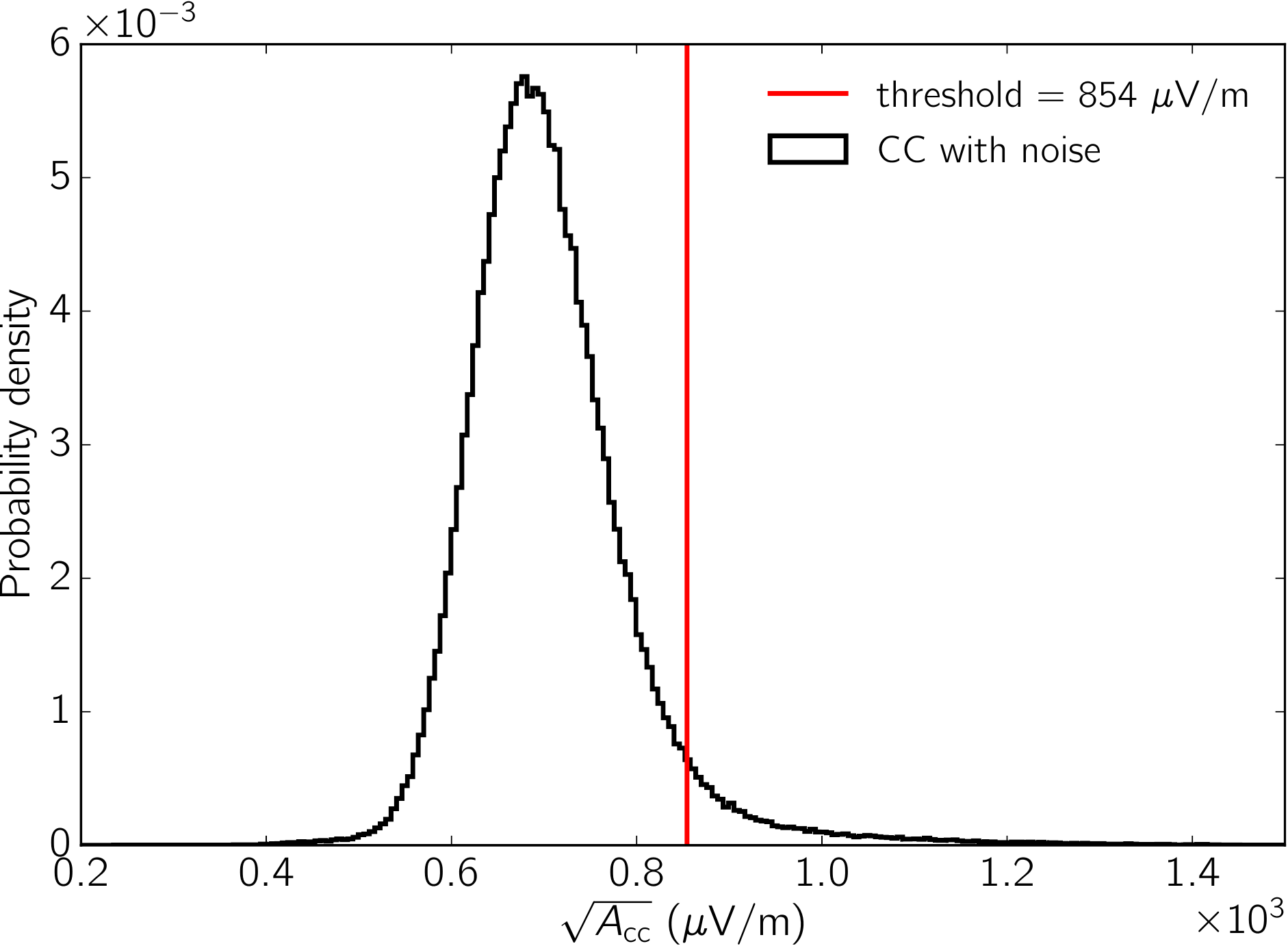}~
	\includegraphics[width=0.48\linewidth]{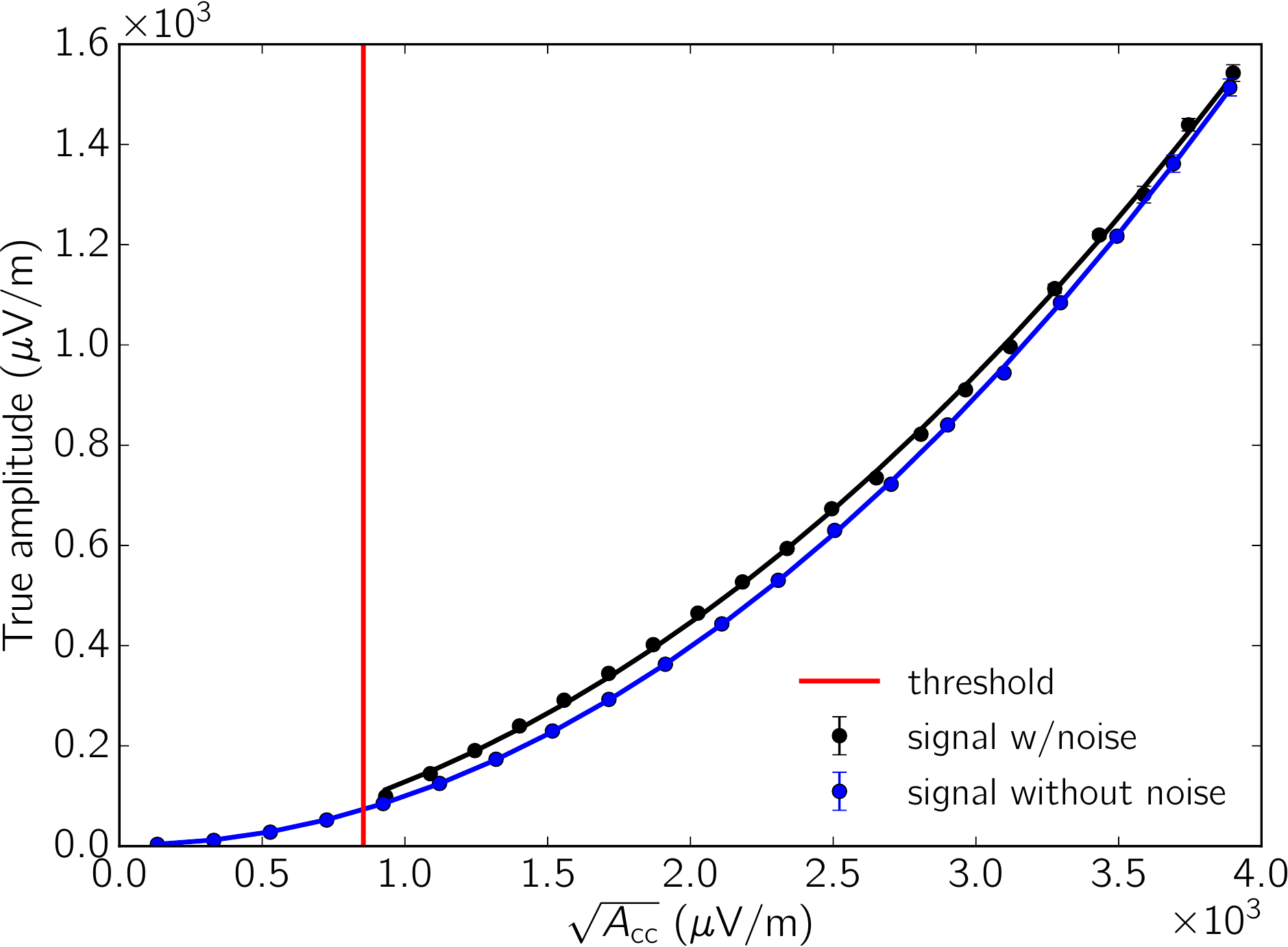}
	\caption{\textit{Left:} Distribution of MF values $\sqrt{A_{cc}}$ applied on traces of pure noise without air-shower signal. 
	The detection threshold (red line) is set to have $<5$\% of false positive detections.
	\textit{Right:} Correlation between $\sqrt{A_{cc}}$ and the amplitude of the simulated signal.}
	\label{fig:mf_thres}
\end{figure}

For testing the filter we use a single template with a length of 60~ns obtained from averaging over many CoREAS simulations.
The threshold of signal detection is set to ${\sqrt{A_{cc}} = 854\,\,\mu}$V/m (see \figref{mf_thres}, left), allowing for a 5\% probability of false positives (similar to our standard method \cite{Bezyazeekov:2015rpa}). 

\subsection{Autoencoder}

The AE~\cite{Bengio:2013} is an unsupervised convolutional network used for learning the coded representation of the data and removing specific features from it.
We train the AE in order to learn noise-related features and clean the input traces leaving cosmic-ray signals.
The input array for our AE consists of 4096 values, what corresponds to a trace length of 1280~ns and 0.3125~ns sampling in order to contain the signal window of 200~ns as well as surrounding background.
For the minimization of the loss, we normalize the input data to the [0:1] range with a baseline level at 0.5.

\begin{figure}[t]
	\centering
	\begin{center}
		\includegraphics[height=0.3\textheight]{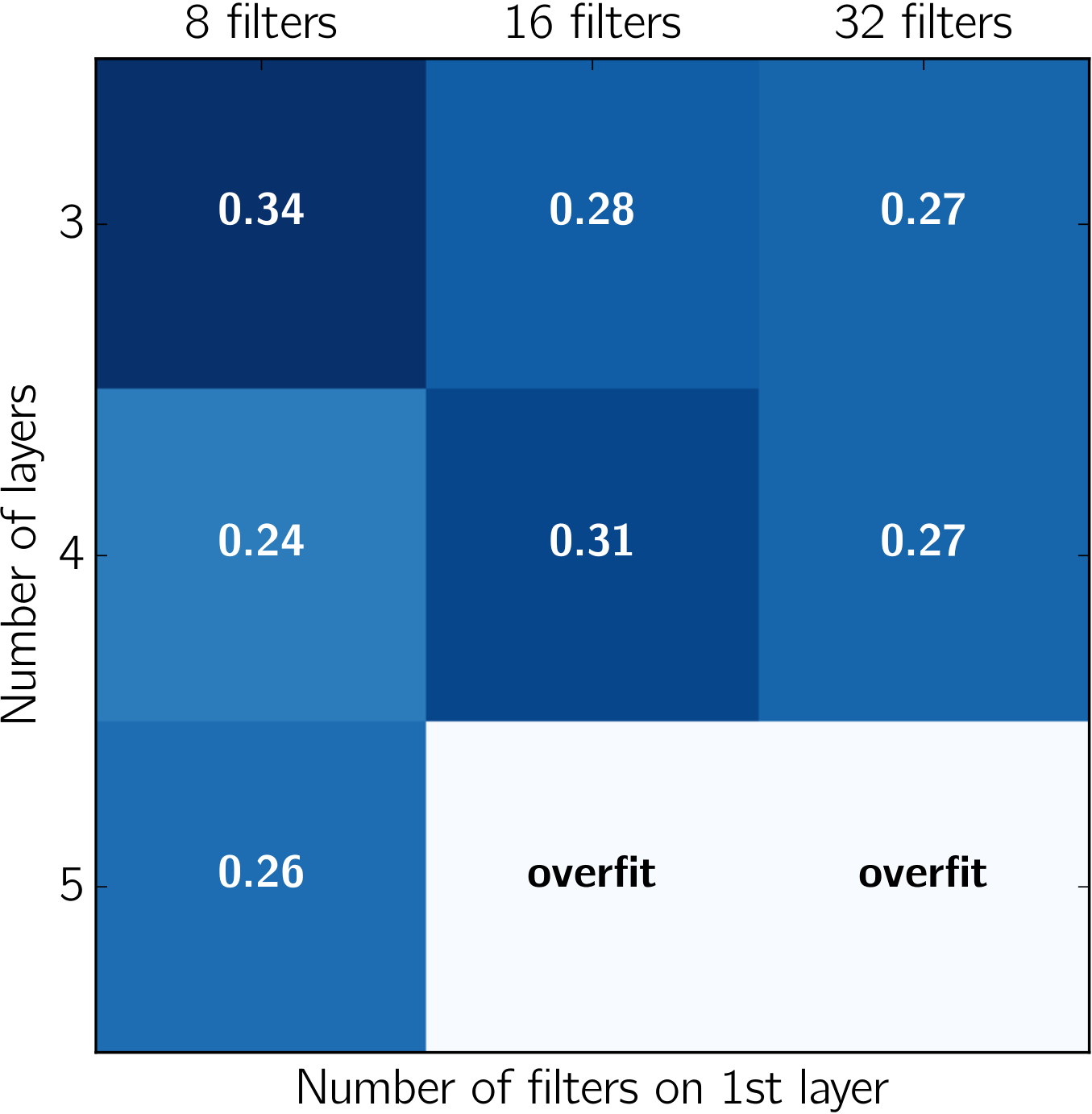}~~\includegraphics[height=0.3\textheight]{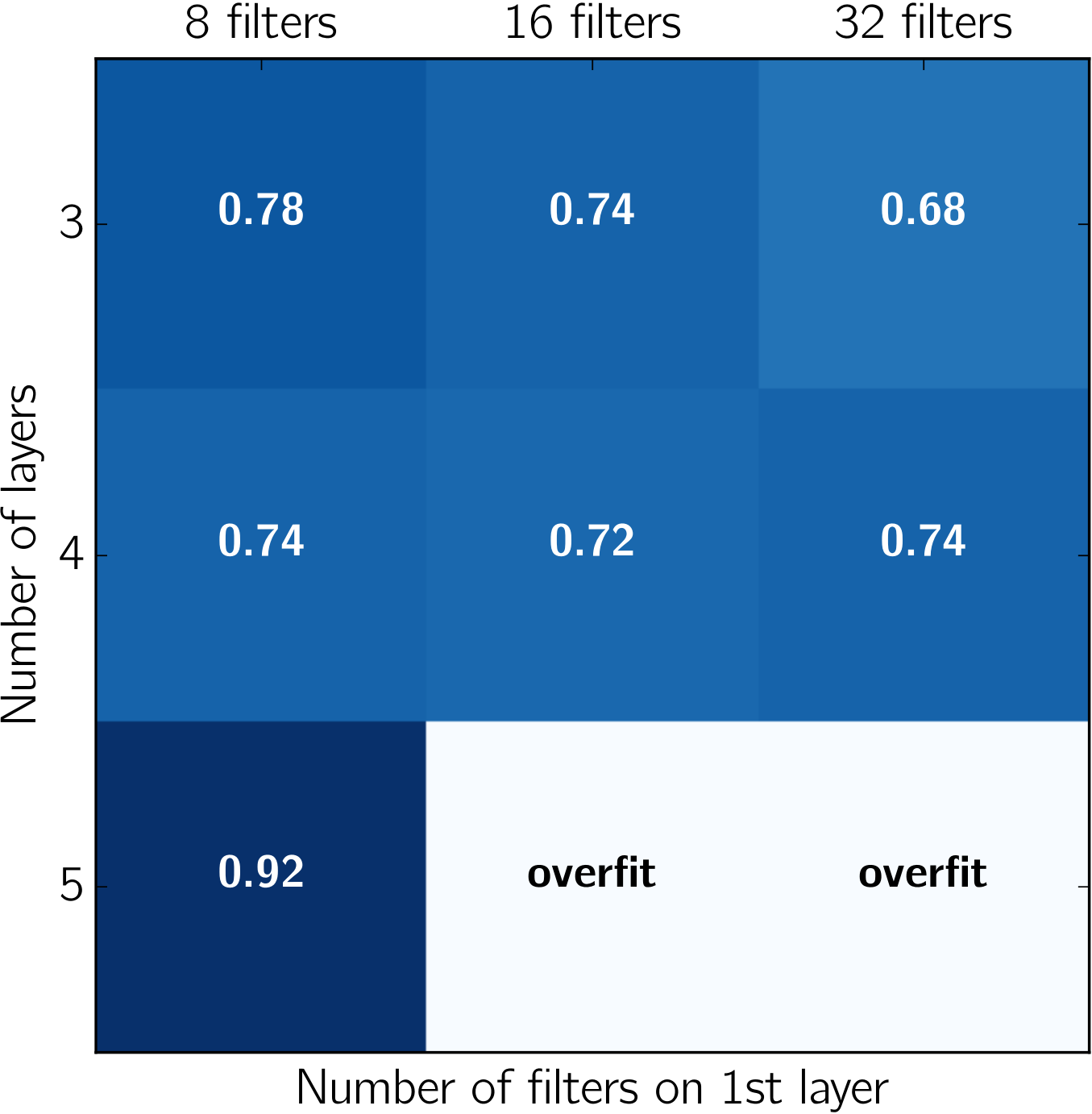}
		\caption{Performance of different configurations of the autoencoder (see main text for explanation). \textit{Left:} efficiency (fraction of reconstructed events $N_{rec}/N_{tot}$). 
		The relatively low efficiency is due to the detection threshold of the AE at the very low SNRs.
		\textit{Right:} purity (fraction of events with properly reconstructed peak position, namely $<5$~ns from true value: $N_{hit}/N_{rec}$).
	    The reconstruction quality of the AE increases with complexity.
		However, the AE has been overfitted when the number of degrees of freedom exceeded the size of the training data.}
	\label{fig:matrix} 
	\end{center}
\end{figure}

The \textit{encoder}, the encoding part of the AE, distinguishes features of noise contained in the input data by applying a set of filters.
The filters perform the convolution of characteristic noise-related features with the input data, estimate its contribution as a result of the convolution, and afterwards send it to the max-pooling layer.
The max-pooling layer performs a discrete downsampling of the data and sends it to the next convolution layer with the next set of filters.
With each layer of the encoder, the data becomes more abstract and reduced in size.
The result of encoding is a map of features of the input data.

\begin{figure}[t!]
	\centering
	\includegraphics[height=0.28\textheight]{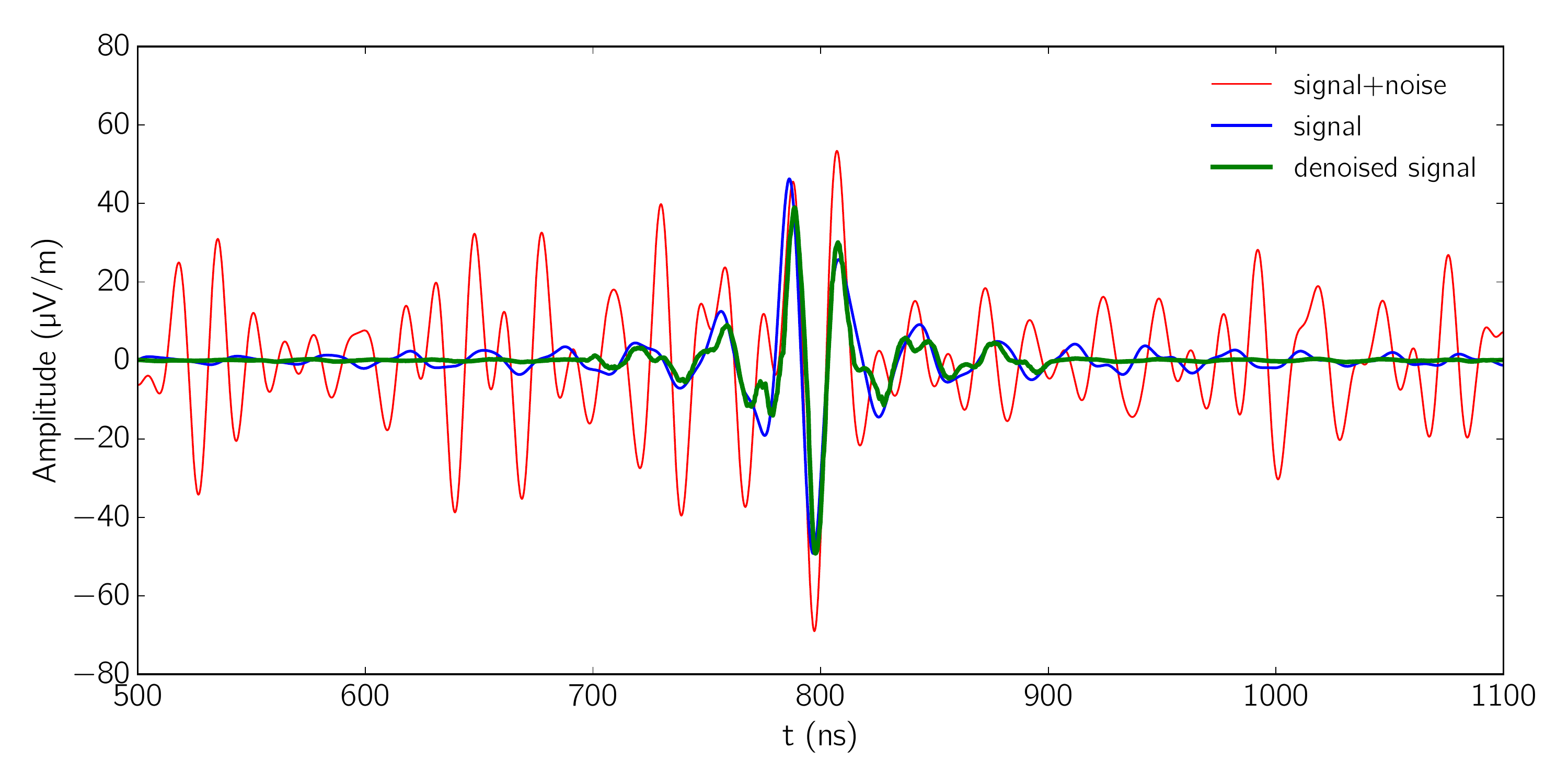}
	\includegraphics[height=0.28\textheight]{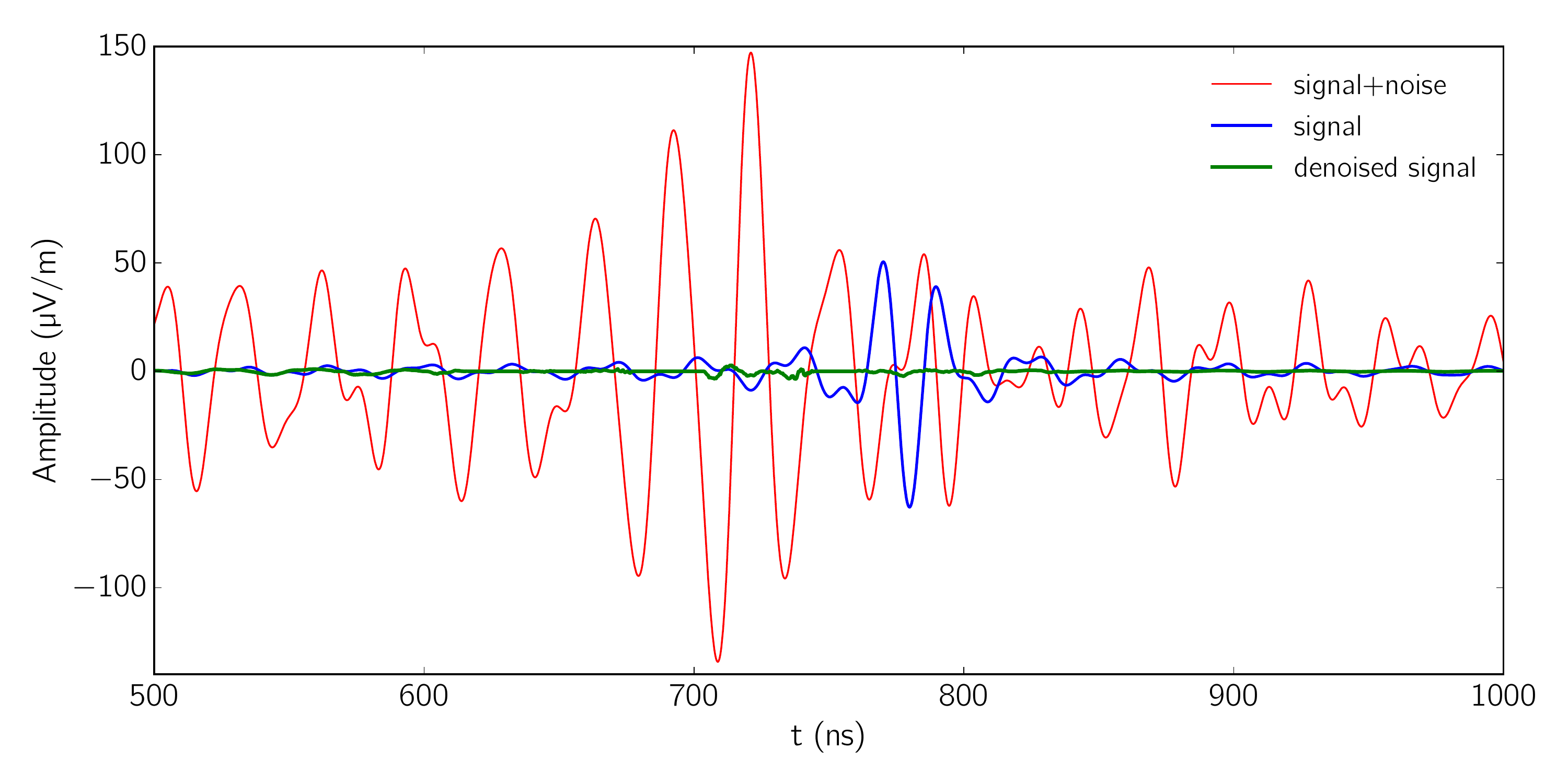}
	\includegraphics[height=0.28\textheight]{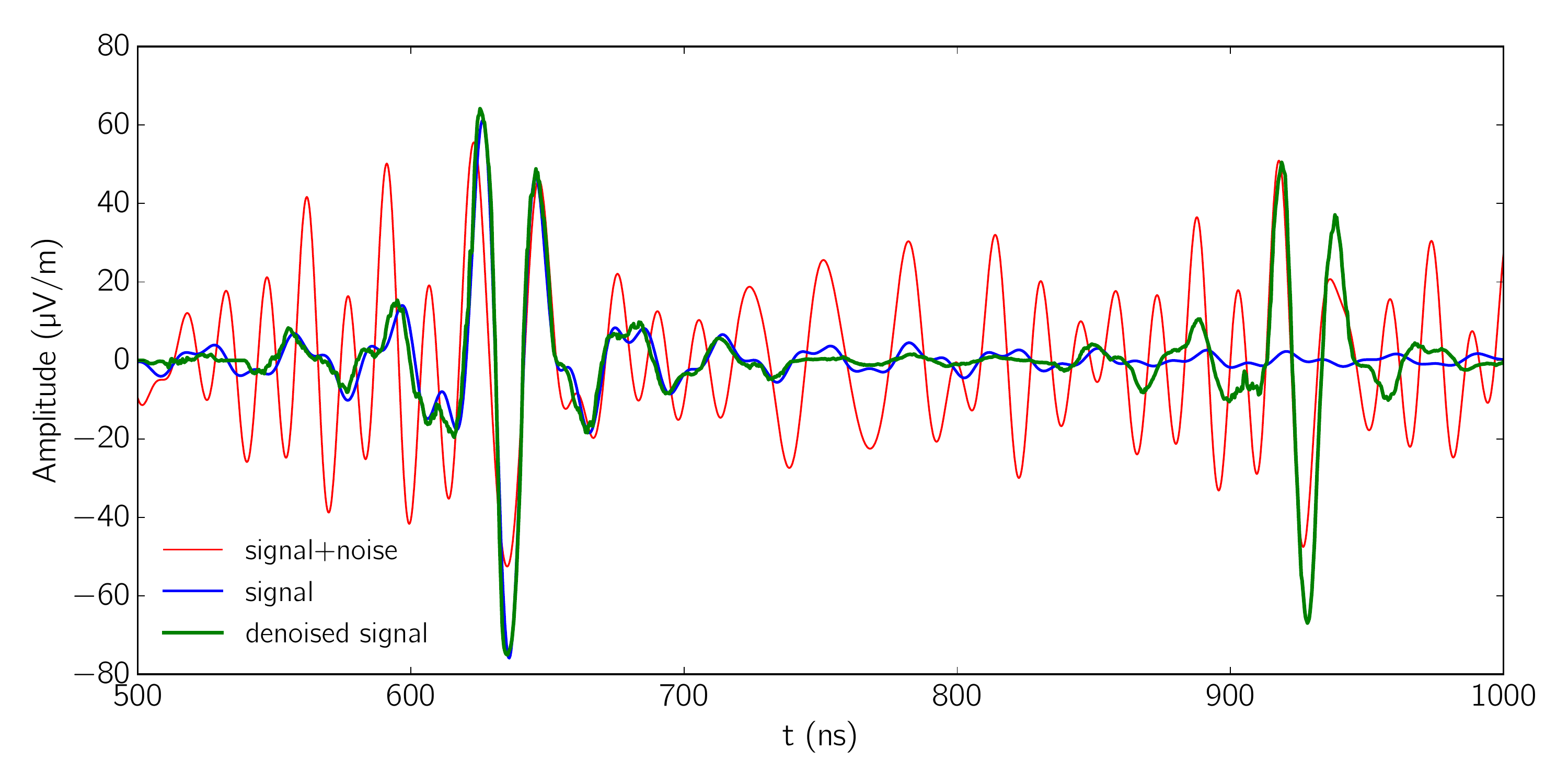}		
	\caption{Examples of the AE performance on simulated data (from top to bottom):
		1) correct identification (true positive, signal is perfectly denoised);
		2) no identification (false negative, due to heavy distortion by the noise),
		3) double identification (true+false positive, passing signal-like RFI).
	}
	\label{fig:reco} 
\end{figure}

After encoding, noise-related features are removed and the map of denoised data proceeds to the next part of the AE, the \textit{decoder}, which produces a reverse reconstruction and returns a data array of the same dimension as the input.
In case of success, the resulting output is the denoised trace containing only the air-shower pulse.

We have explicitly selected a subsample with low amplitudes and low SNR for training to test the possibility of lowering the threshold.
We implemented and trained our AE with Keras~\cite{chollet2015} and Tensorflow~\cite{tensorflow2015-whitepaper} in a uDocker container with GPU support.

We have tested several AE with different sizes defined by the depth ($D$) and number of filters per layer ($N$),
where the configuration of the $i$-th ($i=1,...,D$) encoding layer is defined as follows:
\begin{equation}
S_i = S_{\mathrm{min}} \times 2^{D-i}\,,\,\,\,n_i = 2^{i + N-1}\,,
\end{equation}
where $S_i$ is the size of the $i$-th filter, $n_i$ is the number of filters per layer;
$S_\mathrm{min} = 16$ is the minimum size of a layer. 
To estimate the performance of the AE we use a cross-entropy loss function with the following metrics:
\begin{itemize}
\item \textit{efficiency} $N_\mathrm{rec.}/N_\mathrm{tot.}$, the fraction of reconstructed events. This metric indicates how many non-zero traces are returned by the AE. However this metric alone is insufficient. To evaluate the performance of the AE one has to check also the purity.
\item \textit{purity} $N_\mathrm{hit}/N_\mathrm{rec.}$, the fraction of events with a peak position reconstructed within 5~ns from the true position. This metric indicates the true positive detections and filters misreconstructed signals.
\end{itemize}

\begin{figure}[t!]
	\centering
	\includegraphics[height=0.31\textheight]{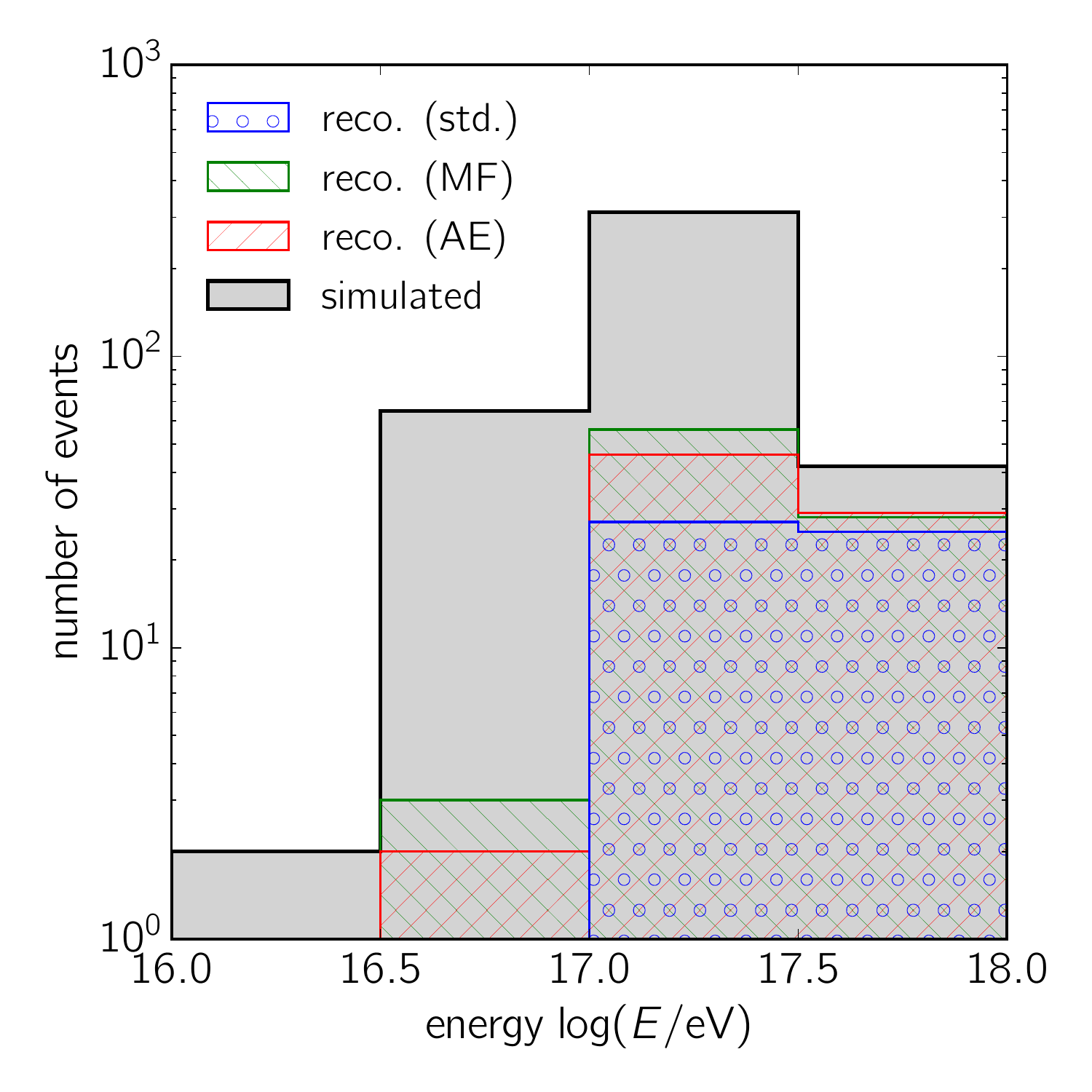}~~\includegraphics[height=0.31\textheight]{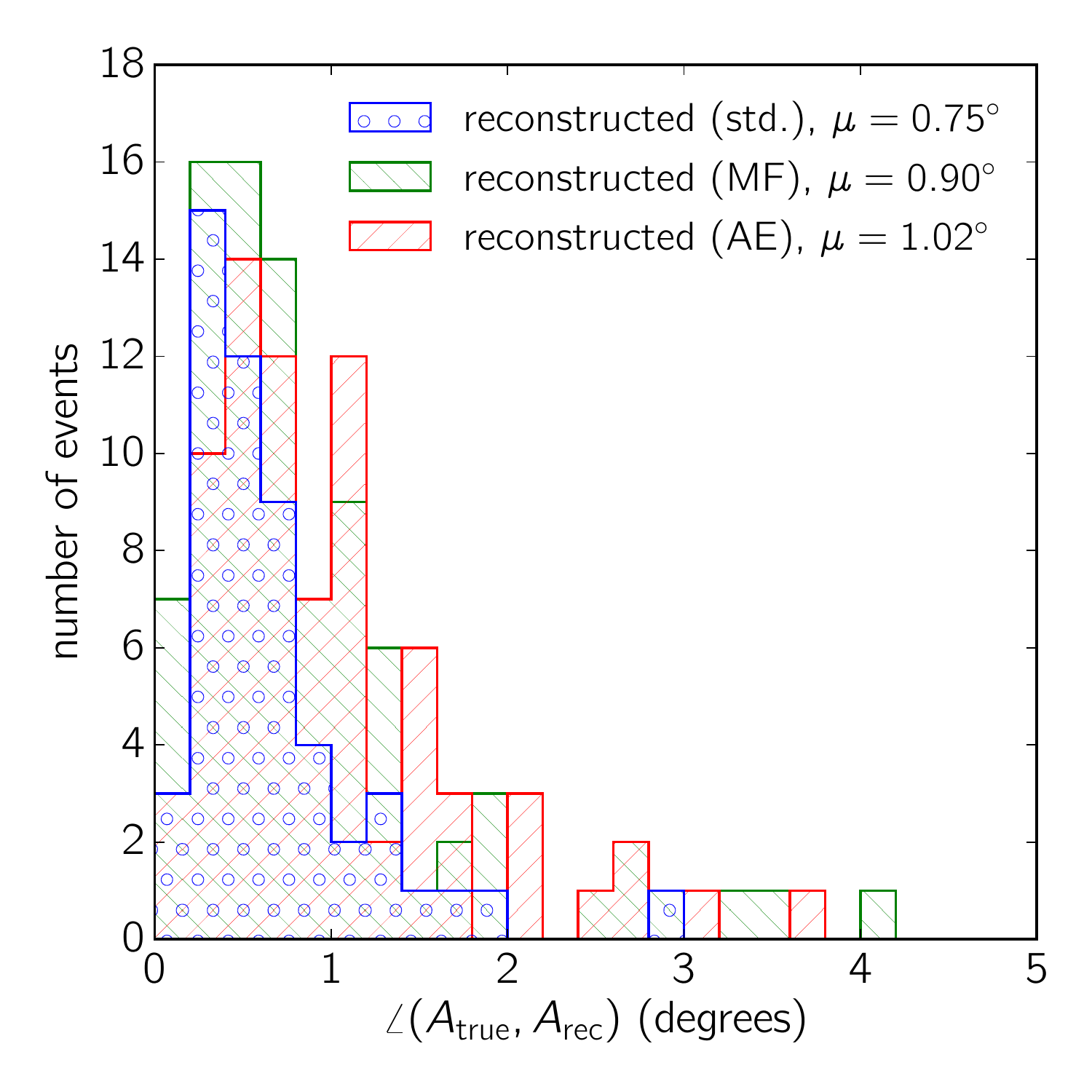}
	\caption{Comparison of the performance of different methods on simulated data. Left: distribution of the reconstructed events as function of the energy for the standard, MF and AE pipelines; right: angular resolution using the different methods.}
	\label{fig:res}
\end{figure}

\figref{matrix} shows a comparison chart of the different AE settings used.
Although only a quarter of the testing sample is reconstructed (due to low SNR), the purity of the reconstruction is very high, i.e. the reconstruction of background-dominated signals is improved.
Moreover, the performance of the AE increases with the complexity of the network and we plan to train more sophisticated AE with larger datasets (our present AE is limited by the relatively small training set).
Examples of the reconstruction are given in \figref{reco}.

\subsection{Comparison of the methods}

\begin{figure}[t]
	\begin{center}
		\includegraphics[width=1.0\linewidth]{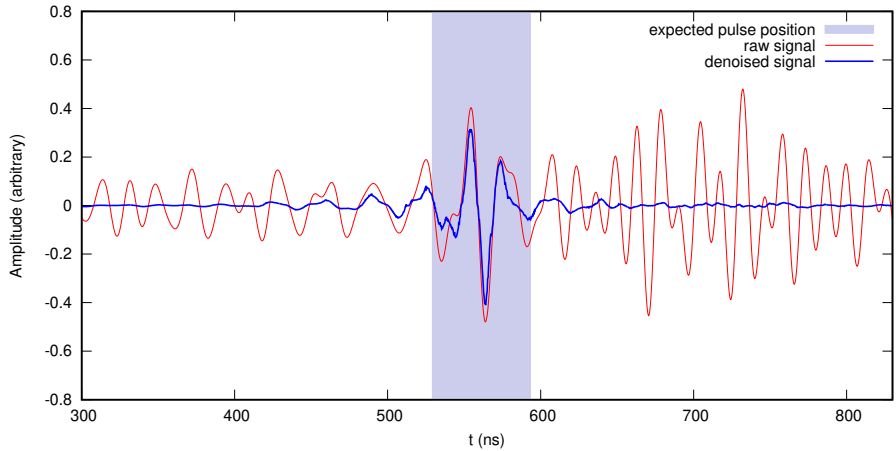}
		\caption{
			Example of the autoencoder performance on a measured Tunka-Rex trace showing successful denoising of the typical RFI after the signal, cf. \figref{nonwhite}.
		}
		\label{fig:real}
	\end{center}
\end{figure}


We have integrated both, the MF and AE algorithms, with the Auger Offline framework~\cite{Abreu:2011fb}, used for the standard reconstruction of Tunka-Rex.
This enables us to check the performance of MF and AE integrated in Tunka-Rex reconstruction pipeline, and to use the identical benchmark for comparison between the standard signal reconstruction, matched filtering, and denoising with the autoencoder.
As metrics we selected the reconstruction of air-shower events (which requires the reconstruction of the signal at several antenna station per event), and the accuracy of the air-shower reconstruction.
We performed this test on simulations using a small subset of the Tunka-Rex library.
The results of the comparison in \figref{res} show that both, MF and AE, are ready for the application to the reconstruction of real data.
Moreover, the threshold is slightly lowered.
With future improvements of MF (using a library of templates instead of a single one, and better whitening of the noise) and the AE (developing a more sophisticated network using a larger training dataset) we expect a further improvement of the reconstruction.
At the moment we are working on the application of the AE on real Tunka-Rex data, an example is shown in \figref{real}.

\section{Discussion and conclusion}
The signal reconstruction in the Tunka-Rex experiment is continuously improved (see, for example, Refs.~\cite{Bezyazeekov:2017jng,Kostunin:2017rbf}).
One of the most advanced method, namely deep learning, represented by the autoencoder technique is discussed in this work (it is worth noticing that a similar technique was recently used for a similar problem~\cite{Erdmann:2019nie}).
We have shown that the methods described in this paper are ready for the implementation for the reconstruction of real data, and we are currently working on this reconstruction.
Besides the application of these methods to Tunka-Rex data we plan to apply them in a Tunka-Rex child engineering experiment, Tunka-21cm, for RFI tagging and denoising, and to the Tien-Shan array~\cite{Beisenova:2017knp} recently upgraded with SALLA.

\section*{Acknowledgements}
This work has been supported 
by the Russian Foundation for Basic Research (grants \textnumero 18-32-00460 and \textnumero18-32-20220),
by the Helmholtz grant HRSF-0027,
and by grant \textnumero 18-41-06003 Russian Science Foundation (Section~2.3).

\bibliographystyle{ieeetr}
\bibliography{references}

\end{document}